\documentclass[%
 reprint,nofootinbib,
 amsmath,amssymb,
 aps,
 pra,
 prb,
]{revtex4-2}
\pdfoutput=1

\newsavebox{\twosubbox}
\usepackage{graphicx}
\usepackage{float}
\usepackage[caption = false]{subfig}

\usepackage{dcolumn}
\usepackage{bm}
\usepackage{xcolor}
\usepackage{url}
\usepackage{hyperref}
\usepackage{parskip}
\hypersetup{
    colorlinks=true,
    linkcolor=red,
    citecolor=blue,      
    urlcolor=cyan,
    pdfpagemode=FullScreen,
    breaklinks=true
    }
\usepackage{amsmath}
\usepackage{amssymb}
\usepackage{physics}
\usepackage{xcolor}
\usepackage{tikz}
\def\be{\begin{equation}}
\def\ee{\end{equation}}
\def\bea{\begin{eqnarray}}
\def\eea{\end{eqnarray}}
\def\Li{{\text{Li}}}
\def\eff{{\text{eff}}}

\begin{document}

\title{Entanglement Entropy in Ground States of Long-Range Fermionic Systems}
\author{Debarghya Chakraborty}
\email{debarghya.chakraborty@uky.edu}
\author{Nikolaos Angelinos}%
 \email{nan236@uky.edu}
\affiliation{%
 Department of Physics and Astronomy,
University of Kentucky, 
Lexington, KY, USA 40506
}%

\date{\today}

\begin{abstract}
We study the scaling of ground state entanglement entropy of various free fermionic models on one dimensional lattices, where the hopping and pairing terms decay as a power law. We seek to understand the scaling of entanglement entropy in generic models as the exponent of the power law $\alpha$ is  varied. We ask if there exists a common $\alpha_{c}$ across different systems governing the transition to area law scaling found in local systems. We explore several examples numerically and argue that when applicable, the scaling of entanglement entropy in long-range models is constrained by predictions from the low-energy theory.  In contrast, disordered models and models without a continuum limit show fractal scaling of entanglement approaching volume-law behavior as $\alpha$ approaches zero. These general features are expected to persist on turning on interactions.

\end{abstract}

\maketitle


\section{\label{sec:level1}Introduction\protect }

Locality severely constrains the features of commonly studied physical systems. Such local systems show special features that are absent in generic quantum-mechanical models. A crucial property is the relative suppression of long-range correlations in the ground states of local Hamiltonians \cite{Eisert_2010} compared to random states in the Hilbert space \cite{Page_1993}.
Besides exhibiting exponential decay of correlation functions, such  ground states obey an area law for entanglement entropy along spatial bi-partitions, in the presence of an energy gap. For $d=1$ critical systems, entanglement entropy may be enhanced by a logarithmic term, whose coefficient is proportional to the central charge of the conformal field theory (CFT) describing the critical point. Similarly, in $d=2$ topological phases of matter can be understood in terms of the presence of certain universal terms in the entanglement entropy \cite{Kitaev_2006}. Entanglement entropy also plays a central role in providing quantum mechanical interpretations for geometric data in theories of quantum gravity \cite{Rangamani_2017}.  
Entanglement entropy has been a
powerful tool to characterize phases of matter and their low-temperature physics. These properties of entanglement entropy have been exploited to efficiently study ground states of many-body systems with the help of various tensor network methods \cite{Cirac_2021}.

In this work, we study the geometric scaling of ground-state entanglement entropy as a function of a continuous parameter which controls the degree of non-locality of interactions. We do this by considering various setups involving fermions with long-range  power-law couplings that decay with the exponent $\alpha$. There are a number of motivations for considering such a setup. Experimental progress has led to the possibility of realizing controlled Ising-type power-law interactions in trapped ion systems with tunable exponent $\alpha$. Such systems carry several signatures of exotic phases of matter, partially reflected by anomalous entanglement scaling.  An example of this is the known logarithmic violation of the area-law behavior in the presence of a gap \cite{Koffel_2012}, \cite{Vodola_2014}.      

The theoretical analysis of long-range models has independent interest. Long-range interactions often show up in continuum theories upon partially integrating out degrees of freedom in local models. While the vacua of such models, in terms of appropriate degrees of freedom, are not expected to be qualitatively different, there are field theories with nonlocal UV descriptions that show novel structures, such as critical points with conformal symmetry without the existence of a local stress-energy tensor \cite{Paulos_2016}, and unusual symmetry-breaking patterns \cite{Chai_2022}. In the context of holography, highly non-local quantum theories with volume-law scaling of entanglement entropy have been proposed as candidate duals to asymptotically flat theories of gravity \cite{Li_2011}.           

In \cite{Kuwahara_2020} it was shown that there exists an area law for general gapped systems having bounded local Hilbert spaces with few-body interactions falling off as a power $\alpha > 2$ in $d=1$ spatial dimensions generalizing previously known bounds for local models \cite{Hastings_2007}, \cite{Arad_2013}. For fermions with long-range hopping or pairing, the bound is tighter and an area law is expected for $\alpha > \frac{3}{2}$ in the presence of a gap. We are interested in understanding the transition to conventional scaling of entanglement entropy as $\alpha$ is varied. The conventional scaling in $d=1$ is an area law (i.e bounded entanglement on increasing subsystem size) for gapped systems and possible logarithmic dependence on subsystem size for gapless systems. There are constructions designed to break these conventional expectations \cite{Vitagliano_2010}, \cite{Samos2}. We ask the following questions: 
\begin{enumerate}
\item For sufficiently small $\alpha$, what kind of scaling of EE do generic models show? 
\item Is there any \textit{universality} in the  transition of EE to conventional scaling as a function of $\alpha$? In other words, does there exist a common $\alpha_{c}$ controlling the behavior of ground state correlations across a variety of models?
\item  How does EE scale for models with a well-defined continuum limit?       
\end{enumerate}

We study these questions with several numerical calculations in quadratic models of spinless fermions on a lattice in $d=1$. Despite the simplicity of these models, their detailed study in the local case has paved the way for understanding interacting and higher-dimensional systems. The numerical flexibility to study large subsystem sizes is a bonus for examining subtleties in the scaling behavior.     

We postulate: 
\begin{enumerate}
	\item Systems with a smooth IR continuum limit will have their long-range entanglement scaling constrained by the scaling of entanglement in their continuum theory. For this reason, such systems will typically have only logarithmic violations of area-law behavior in the presence of a gap for the few-body interactions we consider.
	\item Systems that have large gradients at the microscopic scale, like disordered systems will violate the area law by a power-law correction that may transition to a volume-law at small enough $\alpha$.
	\item The exponent $\alpha_{c}$ at which the transition to conventional scaling occurs depends on features of the system. One may identify $\alpha_{c}$ for particular ensembles of random Hamiltonians. For translationally invariant models in the infinite system limit technical considerations and examples suggest $\alpha_c=1$ though for $\alpha > 1$ the saturation of EE could set in for very large subsystem sizes.
\end{enumerate}

We illustrate all three of the points using numerical examples in free systems that are expected to generalize to the interacting systems. We also support point 1) using some qualitative RG arguments that we hope to make more precise in the future. Section \ref{sec:level2}  deals with models with particle number conservation and discusses the translationally invariant \ref{TImodel} and disordered cases \ref{disorder} separately. Section \ref{pairing} considers translationally invariant models with long-range hopping and pairing. Section \ref{discussion} summarizes the key points along with a conceptual discussion.   


\section{\label{sec:level2}  Models with particle number conservation}
We begin by reviewing some preliminary definitions. The reduced density matrix $\rho_A$ of a contiguous spatial subsystem $A$ with linear size $L_{A}$ is obtained from tracing out the complement $\bar{A}$ from the global state $\rho$ of a lattice of fermion of length $L$ as $\rho_{A} = \Tr_{\bar{A}}{\rho}$. The von Neumann entropy of $\rho_A$ is defined as: 

\begin{equation}
    S(\rho_A ) = - \Tr(\rho_A \log(\rho_A)) 
\end{equation}

When $\rho$ is a pure state, $S(\rho_A)$ is a genuine measure of quantum entanglement and is called the entanglement entropy. The entanglement entropy of eigenstates of lattice models quadratic in fermionic operators can be efficiently computed because the subsystem is specified entirely by two-point functions \cite{Peschel_2003}. The reduced density matrix of a subsystem is proportional to the exponential of a modular Hamiltonian quadratic in fermionic operators. The lattice fermionic degrees of freedom are identified with creation operators $c_{i}^\dagger$. We consider Hamiltonians of the form:

\be
H = \sum_{i, j} V_{ i j} c^{\dagger}_{i} c_{j}
\ee

Here $V_{i j}$ is the Hermitian matrix with entries such that $V_{i j}$ falls of as $ \abs{i - j}^{-\alpha}$ at long distances. This Hamiltonian is diagonalized with a unitary rotation of the operators $c_i$. We adopt the convention that the many-body ground state $\ket{\Omega}$ is the state occupied by the negative-energy modes of the Hamiltonian, without fixing particle number. This means that once we diagonalize $V_{ij}$ as follows,

\be
H = \sum_{i, j} V_{ i j} c^{\dagger}_{i} c_{j} = \sum^{L}_{k = 1} \lambda_{k} \eta^{\dagger}_{k} \eta_{k}
\ee 
the ground-state is given by
\be
\ket{\Omega} = \prod_{k:\lambda_k<0} \eta^{\dagger}_k \ket{0}. \label{defnVac}
\ee

We restrict attention to cases without degenerate ground states. The entanglement entropy of a subsystem $A$ is computed using the eigenvalues of the correlation matrix $({C_{A}})_{ij} = \bra{\Omega} c^{\dagger}_{i} c_{j} \ket{\Omega}$ with $i,j \in A$ as:

\be
S(\rho_{A} ) = - \Tr(C_{A} \log(C_{A})) - \Tr(( \mathbb I - C_{A} ) \log( \mathbb I -C_{A}))  \label{defnEEforC}
\ee 

It follows that empty ground states will have zero entanglement entropy. 

\subsection{Translationally invariant models}
\label{TImodel}
Translationally invariant models have $V_{i j}$ as a function of $\abs{i - j}$ alone, which means that the matrix $V$ is Toeplitz. We suppose $V_{ij}$ can be written in the form 

\be
V_{i j} = \frac{f(d_{O/P}(i, j)) ) }{g( d_{O/P}(i, j))}, \label{defnV}
\ee 

 where $f(r)$ is a function that remains bounded and does not decay with $r$ and $g(r)$ is a function that grows as $r^{\alpha}$. For finite systems one may consider open boundary conditions $d_{O}(i, j) = \abs{i-j}$, or periodic boundary conditions with distance $d_{P}(i, j) = \min(L - \abs{i -j}, \abs{i-j} )$ such that $V$ is circulant.  
In the thermodynamic limit, the spectrum of the Toeplitz matrix coincides with that of the circulant matrix. The latter can be diagonalized even at finite $L$ using a Fourier unitary. The choice of open or periodic boundary condition does affect the entanglement entropy but not its scaling in the thermodynamic limit,  see  for example \cite{Calabrese_2009}. For finite $L$, the open boundary condition is the physical choice. For $\alpha \leq 1$ in 1D, the spectrum may develop isolated divergences in the thermodynamic limit, related to the nonconvergence of the generalized harmonic sum. The ground-state energy-density may still remain well defined.  


Let us consider the large $L$ limit of these models. Our claim is that for a generic translationally invariant $V_{i j}$ which \textit{also has a smooth continuum limit}, the entanglement entropy is at most $\log(L_{A})$ for any $\alpha$. This  well-understood point for short-ranged models generalizes in a simple way and has important implications. We briefly review the mathematical and conceptual underpinnings of this result. 

In the thermodynamic limit, the eigenvalues become a smooth function of quasimomenta $k$ taking values in $[0, 2 \pi)$, in units of inverse lattice spacing  
\be
\lambda(k) =  \sum^{\infty}_{r=-\infty}  V(r) e^{i k r},
\ee 
where $V(r) = V(i-j)$. The correlation submatrix $C_{A}$ of the subsystem of interest is given by 
\be
 C_{A}(i-j) =  \expval{c^{\dagger}_{i} c_{j}} = \frac{1}{2 \pi} \int^{2 \pi}_{0} dk \quad e^{-i k (i -j)} \Theta(-\lambda(k)) \label{scalarsymbol},   
\ee 
where we used the important fact that $C_{A}$ is Toeplitz to write  $C_{A}(i-j)$. The formula \eqref{defnEEforC} can be rewritten as a contour integral involving  $\log(\det(C_{A} - \mathbb I_{A}))$ and model independent functions. On using the Fisher-Hartwig conjecture and certain assumptions \cite{Jin_2004}, the leading answer is proportional to $\log(L_{A}) $. The prefactor of the $\log$ is given by $\frac{1}{6}$ times the number of discontinuities of the ``symbol'', $\Theta(-\lambda(k))$. These discontinuities are precisely at the Fermi points $k^{*}_{i}$  where the dispersion changes sign. The 1D answer for free fermions was generalized to higher-dimensions in \cite{Gioev_2006} and seen to be consistent with a conjecture due to Widom. 

There is a physical argument based on RG for the results explained above as discussed in \cite{Calabrese_2009}, \cite{Swingle_2010} that we expand on.  The long distance entanglement properties of such free fermion models under consideration should match the predictions from the effective IR theory. Despite the potentially complicated nature of the dispersion $\lambda(k)$, assuming analytic behavior about $c$ Fermi points $k^{*}_{i}$, the IR theory in momentum space will look like: 

\be
H \propto \sum^{c}_{i=1} \int dk \, v_{F_{i}}  k    (\psi^{\dagger}_{i} \psi_{i} -\chi^{\dagger}_{i} \chi_{i} )  + \ldots, \label{effectivele}
\ee
where ${v_{F}}_{i} = \lambda'(k^{*}_{i})$ is the local Fermi velocity and $\psi_{i}$ and $\chi_{i}$ are the right and left-movers corresponding to low energy excitations and holes. We omit the higher powers of $k$, which are irrelevant at low energies. Thus the entanglement entropy can be understood as being the sum of contributions from $c$ decoupled chiral and anti-chiral modes. The CFT answer is $\frac{n_{L} + n_{R}}{6} \log(L_{A})$ which gives a contribution of $\frac{1}{6} \log(L_{A})$ from each mode. On summing them we get $S(\rho_{A}) = \frac{c}{3} \log(L_{A}) + \ldots$ where the ellipsis stands for subleading contributions. The answer for open boundary condition can be derived within the CFT formalism and is given by halving the prefactor of the logarithm $S(\rho_{A}) = \frac{c}{6} \log(L_{A}) + \ldots$.

\begin{figure}[htp]

        \subfloat[]{\includegraphics[width=1\linewidth]{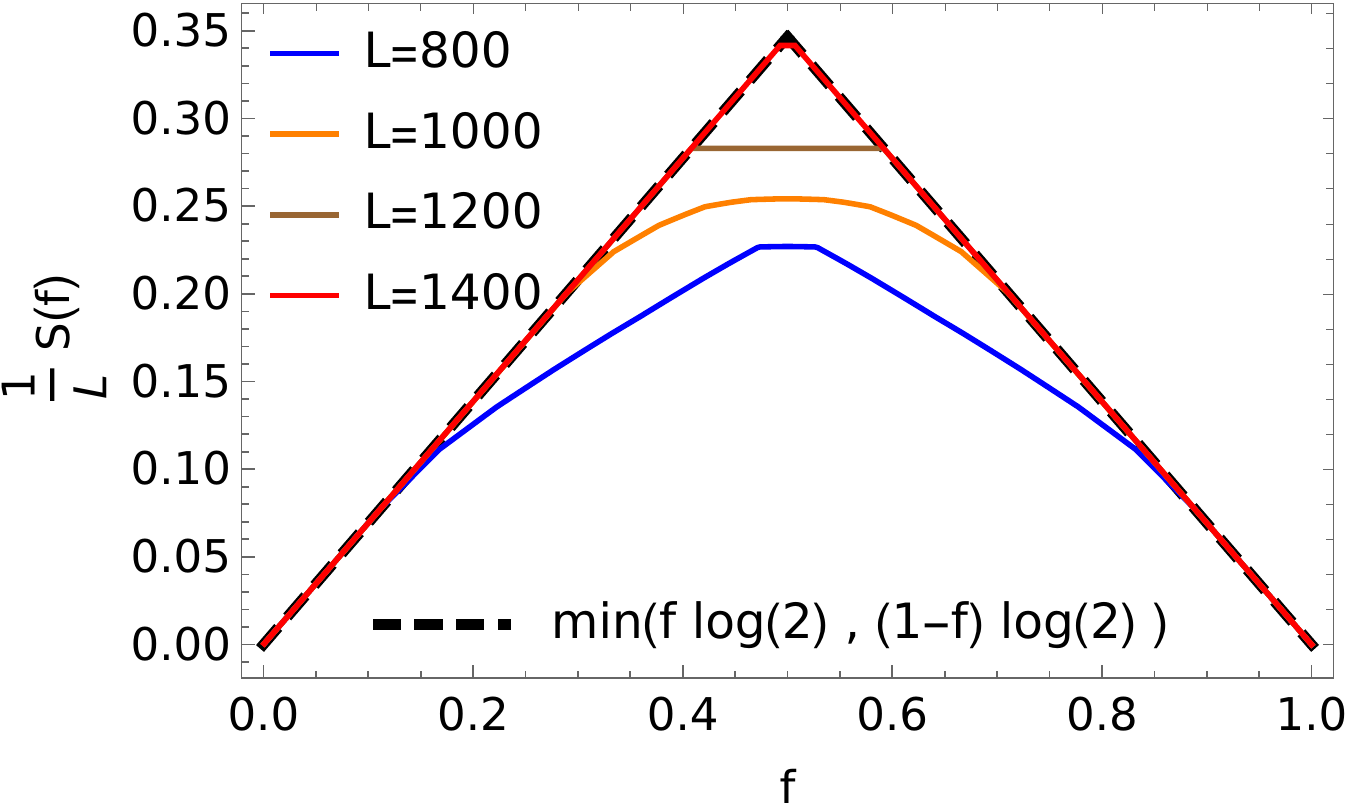}\label{MaxGrowthEE}}
        \\
       \subfloat[]{\includegraphics[width=1\linewidth]{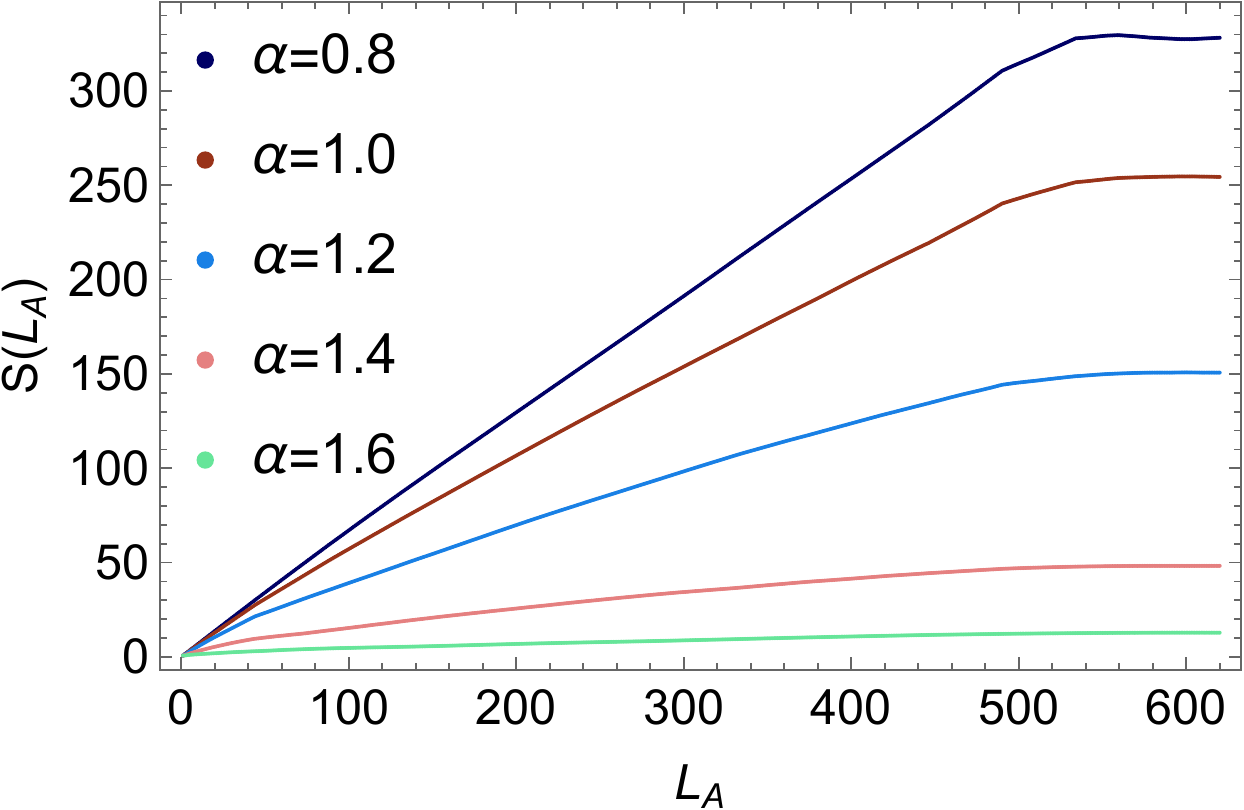}\label{transitionforDetModel}}
       \\
	\subfloat[]{\includegraphics[width=1\linewidth]{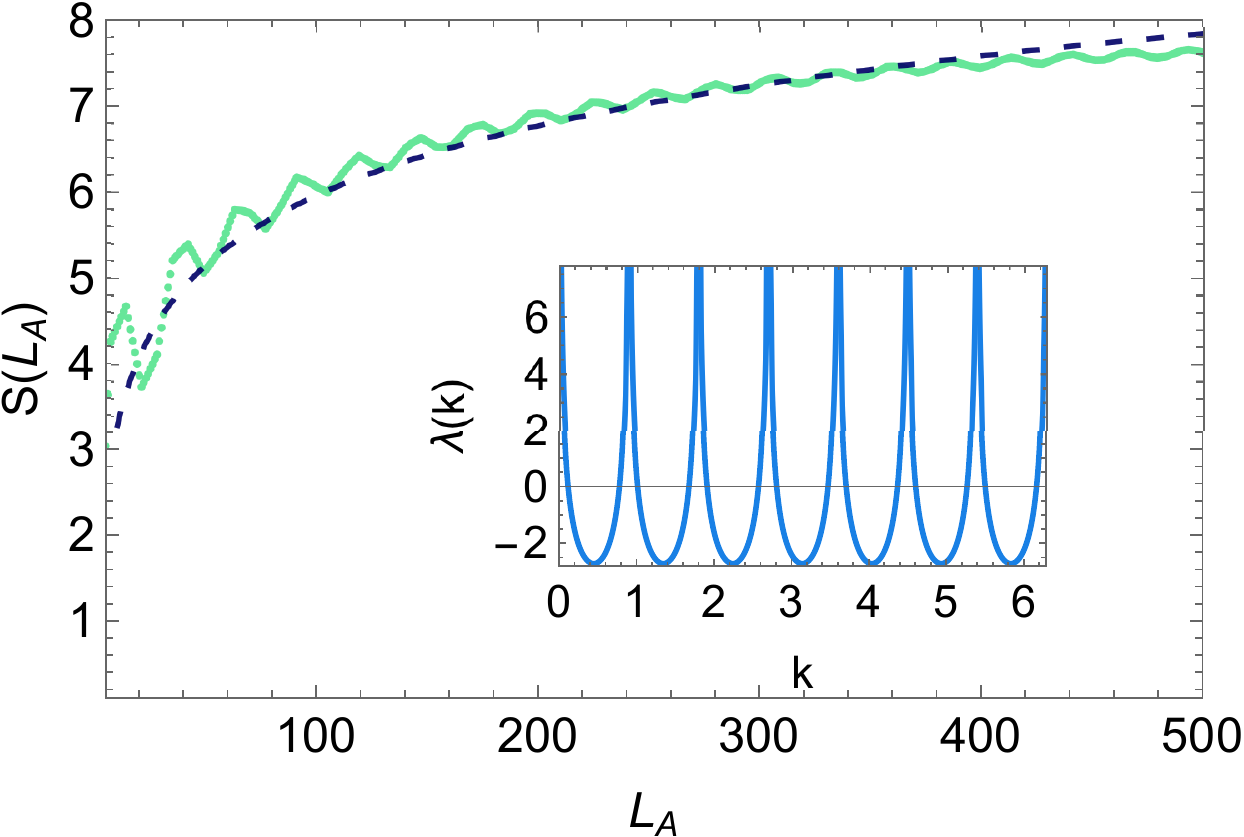}\label{OBC_c}}
    \caption{ Fig \ref{MaxGrowthEE} shows maximal volume-law growth of entanglement in ground state of the model specified by \eqref{maximalee} at $\alpha=0$, shown for sequence of $L$ starting from $800$ to $1400$.  To compare across system sizes, the entropy density $ \frac{1}{L} S $ is expressed as a function of subsystem fraction $f = \frac{L_{A} }{ L }$. Fig \ref{transitionforDetModel} shows the intermediate EE scaling regime for the same model for different $\alpha$ at $L=1200$. Fig \ref{OBC_c} shows the agreement between the EE of a long-ranged hopping model with $\alpha = 0.6$ of $L=1200$ and OBC, with the CFT predictions (dashed lines). The inset shows the dispersion relation of this model for $L\rightarrow \infty$ limit with seven pairs of Fermi points giving rise to $c = 7$ in the CFT formula. }
  \end{figure}

The low-energy theory written in \eqref{effectivele} has an emergent Lorentz invariance but lacks conformal invariance unless all the ${v_{F}}_{i}$ are equal. In that case conformal symmetry arises from a spacetime rescaling, in addition to the internal symmetry between different chiral fields. An interesting manifestation of the emergent conformal symmetry is found in examining the finite-size correction to ground state energy: 
\be
F_{L} =  f_{0} L  - \frac{\pi c v_{F}}{6 L} + O(\frac{1}{L^2} ), 
\ee 
where $f_{0}$ is the ground state energy density in the thermodynamic limit and $F_{L}$ is the ground state energy at finite size $L$. This relation is obtained applying the Euler-Maclaurin formula to an arbitrary dispersion on the lattice. The leading order answer precisely matches the correction to vacuum energy density of a CFT on mapping a theory from infinite plane to a cylinder of radius $L$ and setting $v_{F} = 1$.  
The lack of scale invariance of \eqref{effectivele} is not a problem from the RG perspective, because the $v_{F_i}$'s get renormalized and the theory flows to the one with all $v_{F}$'s the same. This expected behavior is illustrated with the agreement between finite-size results and the thermodynamic limit of 
\be
V_{i j} =\begin{cases}
	 \frac{1}{d_{P}(i,j)^{0.6} } \sum^{l=3}_{l=-3} \cos(\frac{2 \pi l}{7}  d_{P}(i,j) ) & i\neq j\\
	 0 & i=j
\end{cases}.  \label{OBC_Ex}
\ee  
The function $\lambda(k)$ can be computed in the thermodynamic limit as a linear combination of polylogs. The periodic modulation that picks out every seventh site in the numerator gives rise to $c =7$ pairs of Fermi points and emergent species of fermions. The EE for $L=1200$ with open boundary conditions is presented in Fig. \ref{OBC_c}.  This example illustrates how predictions from the effective theory give an accurate answer for the EE when there are no obvious means to an exact answer: for finite $L$ and open boundary conditions.      

The nontrivial implication of the argument in \cite{Swingle_2010} was that it predicted, in $d$ dimensions, the same $L_{A}^{d-1} \log(L_{A})$ scaling of EE for interacting theories that can be described in terms of a deformed Fermi surface. This discussion generalizes to our setting: for fermionic systems with a continuum limit, provided the Fermi surface is stable, the leading piece of EE will continue to scale at most as $L_{A}^{d-1} \log(L_{A})$ even if the microscopic degrees of freedom have long-range interactions. The stability of the Fermi surface is a more subtle matter, but it is unlikely that long-range hopping terms alone affect the conventional kinematical arguments provided at least that $\alpha > d$, though long-range density interactions might. The hopping model ground states considered above are simple and non-generic, but they illustrate the utility of effective theory for understanding entanglement.



Lattice models without a continuum limit however, are not constrained in this manner. By this we mean models which lack a gradient expansion or equivalently, a smooth momentum-space Hamiltonian which can be expanded around low-energy points. Disordered hopping models are known to exhibit volume law entanglement in the limit of small $\alpha$, \cite{Liu_2018}, and are expected to show area law for large $\alpha$. We study this transition in EE  for disordered models in Section \ref{disorder}. 

However, disorder is not necessary to go beyond $\log(L_{A})$ scaling as has been appreciated in \cite{Fannes_2003}, \cite{Farkas_2005}.  
To illustrate this point, we first construct an example of $V_{i j}$ with translation invariance weakly broken only due to open boundary conditions, that we numerically show to saturate the maximal possible growth of entanglement for fermionic models at $\alpha = 0$, see Fig \ref{MaxGrowthEE}. This is given by the highly oscillatory sequence of models: 
\be
V_{i j} = \begin{cases}\frac{1}{d_{O}(i, j)^{\alpha}}\frac{ \sin( (L+\frac{1}{2} ) d_{O}(i,j) ) }{\sin ( \frac{1}{2} d_{O}(i,j) )}  &  i \neq j\\
0 & i=j \end{cases}
 \label{maximalee}
\ee 
The transition to bounded EE is seen in this model through the appearance of a plateau at large fractions $f = \frac{L_{A}}{L}$ and the appearance of fractal scaling $L^{\gamma}_{A}$ with $0 < \gamma < 1$ for $ 0.8 < \alpha< 1.6$, see Fig. \ref{transitionforDetModel}. The intermediate fractal scaling $L^{\gamma}_{A}$ is an interesting feature that shows up robustly for  disordered models. We expand on this in Section \ref{disorder}. Note that this sequence of models has couplings explicitly dependent on $L$ and therefore its scaling properties might differ from conventional models. 


Another example of a translationally invariant model without disorder with a ground state which saturates the maximal growth of entanglement entropy is the following
\be H=\sum_{j=1}^\frac{L}{2} c_{j+{L\over 2}}^\dagger c_{j}+ \sum^{\frac{L}{2}}_{j=1} c_{j}^\dagger c_{\frac{L}{2} + j},\ee
where we assume that the system size $L$ is even. The entanglement entropy for $L_A\leq L/2$ is $S(L_A)=L_A \log 2$ (see appendix \ref{appendix:b}).

\subsection{Universality in Disordered Models}
\label{disorder}

Volume-law scaling of EE in random hopping models has been studied across the entirety of spectrum in \cite{Vidmar_2017},\cite{_yd_ba_2020} and for ground states in \cite{Ranjan}, \cite{Juhasz}, \cite{Liu_2018}, the latter in the context of the SYK$_2$ model. SYK$_q$ models are generally interpreted as being $N$ flavors of fermions embedded in a quantum dot, with all-to-all interaction in flavor space. Here we consider a disordered hopping model that interpolates between a complex SYK$_{2}$ model embedded on a 1D wire at $\alpha \rightarrow 0$ to a disordered local model for $\alpha \rightarrow \infty$. We examine the nature of the transition of EE scaling and the value $\alpha_{c}$ that controls the transition to bounded EE.

For our numerical study we consider the ensemble:
\be
V_{i j} = \frac{R_{i j}}{1 + ( d_{O}(i,j) )^\alpha}, \label{powerlawrm}
\ee 
where $R_{ij}$ are elements of a symmetric Gaussian random matrix with zero mean. The variance of $R_{ij}$ needs to be chosen as a function of both  $L$ and $\alpha$ to retain extensivity of total energy, but the overall scale does not affect the ground state or entanglement properties.     
The single particle Hamiltonian is an example of a power-law banded random matrix of the type studied in \cite{Mirlin_1996}. This model is known to exhibit an Anderson localization transition at $\alpha = 1$, as diagnosed through inverse participation ratios \cite{Mirlin_1996}, \cite{Mirlin_2000}. The random matrix model in \eqref{powerlawrm} retains a lack of correlation between matrix elements akin to Gaussian random matrix models, but loses invariance under a change of basis.   

The ensemble averaged EE $\overline{S}(L_{A})$ displays the following properties: 
\begin{enumerate}
	\item Self-averaging, meaning that a randomly picked member of the ensemble gives $S(L_{A})$ similar to $\overline{S}(L_{A})$ for fixed $\alpha$ and $L$. The distribution of $S(L_{A})$ about the mean increases with $\alpha$.
	\item $L$-independent behavior for $L_{A}/L$ small compared to $\frac{1}{2}$.
	\item Leading order fractal scaling $L_{A}^{\gamma(\alpha)}$ with $0 <\gamma(\alpha) < 1$ with $\gamma(\alpha)$ continuously decreasing from 1 as $\alpha \rightarrow 0$.    
\end{enumerate}      

\begin{figure}[!htp]
        \subfloat[]{\includegraphics[width=1\linewidth]{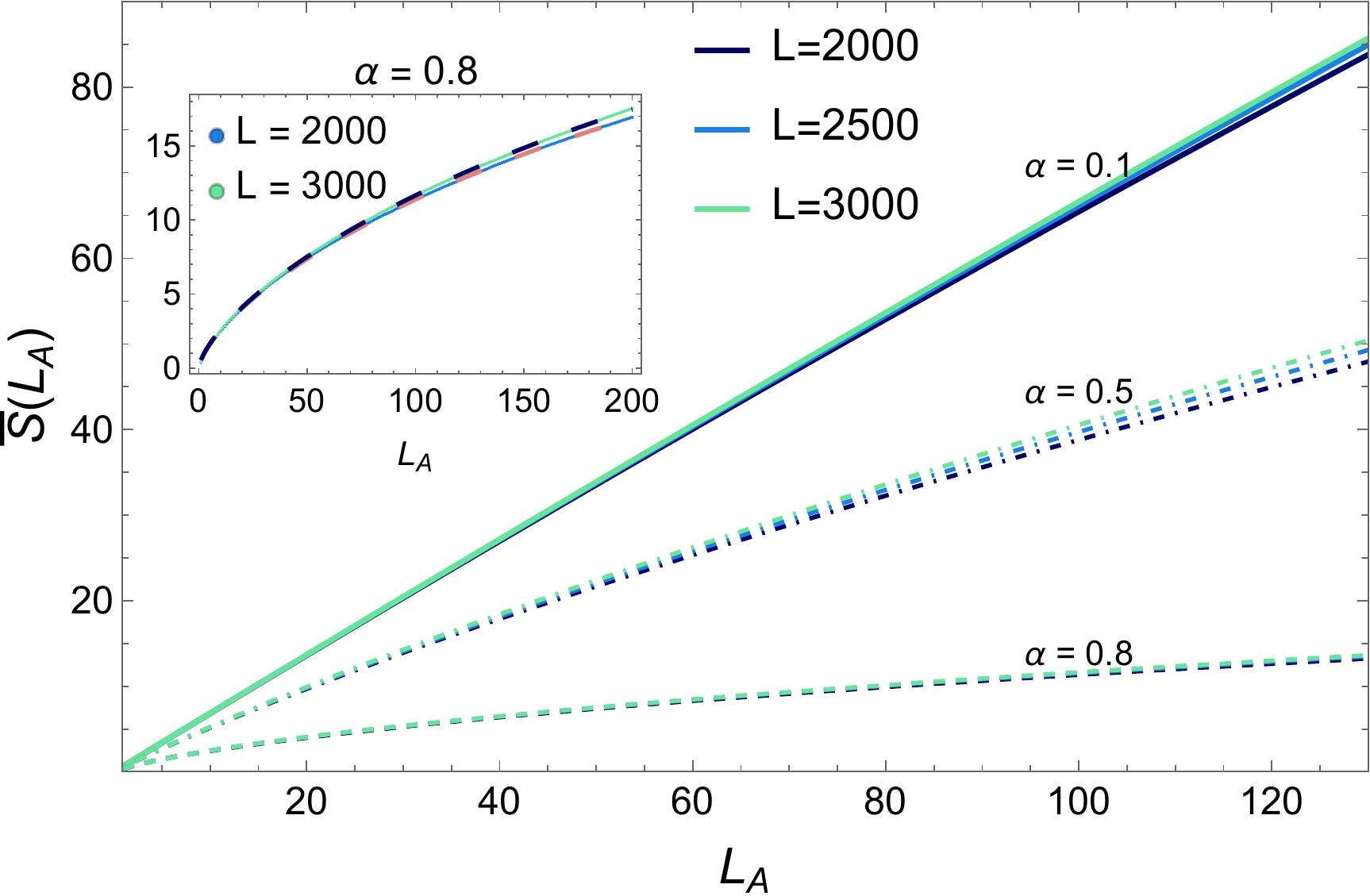}\label{DemoPL}}
        \\
		\subfloat[]{\includegraphics[width=1\linewidth]{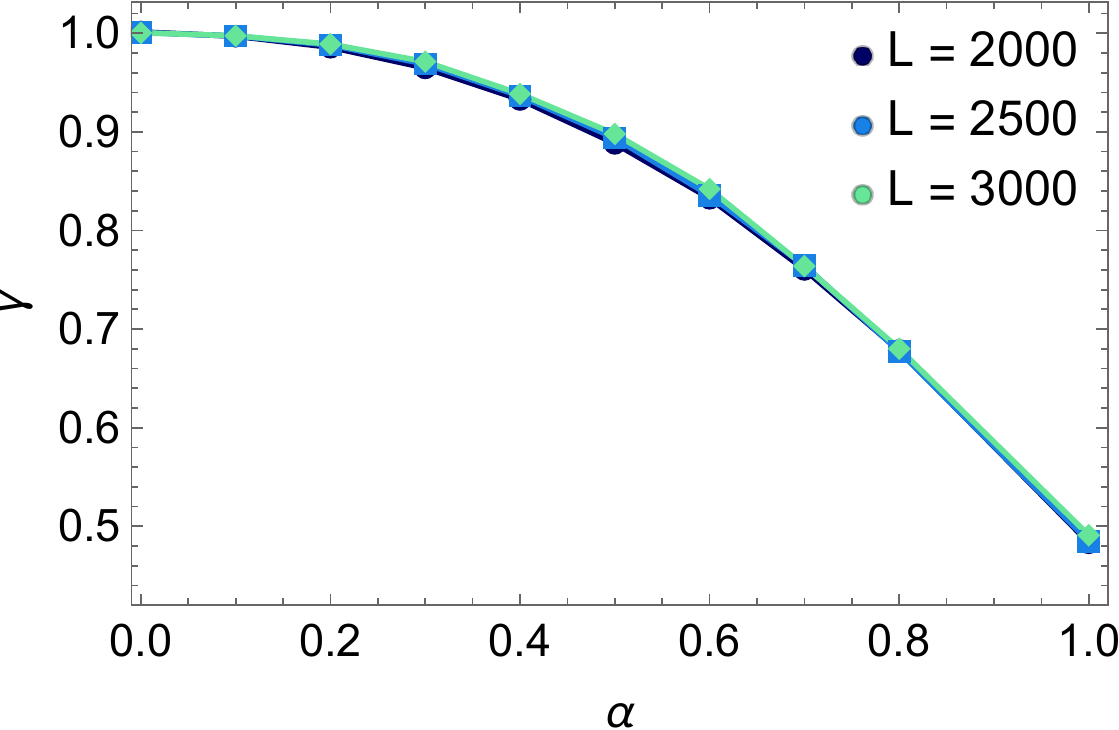}\label{gammaasfunctionalpha}}
		\\
    	\subfloat[]{\includegraphics[width=1\linewidth]{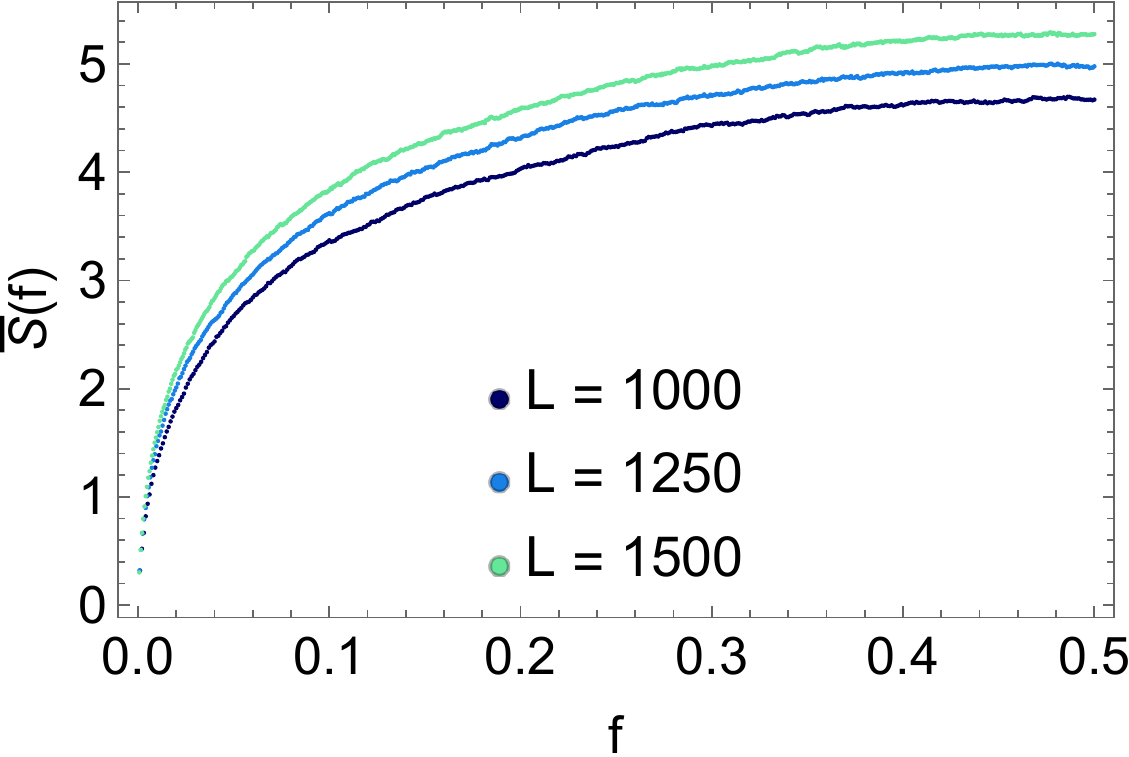}\label{1point05}}
    \caption{Fig \ref{DemoPL} shows the behavior of the ensemble averaged $\overline{S}(L_{A} )$ for different $\alpha$'s and its $L$ independent collapse, for $L_{A}$ smaller than a certain fraction of $L$. Inset shows $\overline{S}(L_{A})$ for $\alpha = 0.8$ and $L=2000, 3000$ along with the fit \eqref{fractalscaling} indicated by dashed lines. Fig \ref{gammaasfunctionalpha} shows numerically obtained best-fit $\gamma(\alpha)$ for system sizes $L=2000, 2500, 3000$. To compare across system sizes best fit parameters were obtained by fitting $\overline{S}$ for $L_{A}$ up to $200$. The collapse for different system sizes over the range of $\alpha$ indicates $\gamma$ for this range may be well-defined in thermodynamic limit for finite $L_{A}$. For $\alpha>1$, extracting the value of $\gamma$ becomes difficult, though the EE still shows growth. Fig \ref{1point05} shows that for numerically accessible $L$, $\overline{S}$ now expressed as function of $f=\frac{L_A}{L}$ continues to slowly increase at $\alpha=1.05$. Between 400 to 500 samples were used to generate the plots above. }     
  \end{figure}

We numerically checked that other choices of random ensembles of $V_{ij}$ preserve the features mentioned above as long as the $V_{i j}$'s are identically distributed and uncorrelated. For example, a sign-randomized ensemble with $R_{ij} = \pm 1$ will show similar features. The exponent $\gamma(\alpha)$ as well as the coefficient of the term $L^{\gamma(\alpha)}_{A}$  are dependent on the choice of ensemble;  see Figure \ref{gammaasfunctionalpha}. 

To account for finite-size effects we use the following fitting function:
\be
\bar{S}(L_{A}) = a L^{\gamma}_{A} + \frac{b}{L} L^{\gamma + 1}_{A}. \label{fractalscaling} 
\ee

This fitting function is chosen for two reasons. First, the Page curve for free fermions is known to be susceptible to considerable finite size effects. In the $\alpha \rightarrow 0$ limit, $\gamma = 1$ and the leading order correction to volume-law behavior is of the form $\frac{L^2_{A}}{L}$ \cite{_yd_ba_2020}.
Second, the best-fit $a(\alpha)$ and $b(\alpha)$ are $O(1)$ numbers roughly independent of system size.  Therefore, at least for $f = \frac{L_{A}}{L} \ll 1$ and $\alpha$ not too large, this fit should capture the correct behavior.

The fit gives accurate behavior, see the inset in Fig \ref{DemoPL}. For small $\alpha$, say $0.01$, we checked that the answer for ground state EE matches the average answer for EE across the spectrum valid for arbitrary subsystem fractions derived in \cite{_yd_ba_2020}. This fractal scaling is consistent with the results of scaling of entanglement for typical eigenstates in a disordered ensemble studied in \cite{Ranjan}. 
The rate of change of $\gamma(\alpha)$ sharply increases at about $\alpha \approx 0.5$, see Fig \ref{gammaasfunctionalpha}. The transition to bounded entanglement at larger $\alpha$ seems to be continuous. That makes it difficult to pin down a precise $\alpha_{c}$ where the ensemble-averaged $\overline{S}$ saturates as for short-ranged disordered hopping models.
We estimate that $\alpha_{c} \approx 1.3$ where according to our numerics, $\overline{S}$ saturates for a sequence of $L$. We note that for the range of powers we consider, there is no indication of an emerging gap. 
Although the localization of entanglement and Anderson localization arise due to similar mechanisms in the considered model, $C_{A}$ and therefore entropy depends on the correlations between distinct single-particle eigenvectors. Therefore without contradiction, the EE continues to show slow fractal growth for $\alpha>1$ when the model has undergone a localization-transition as diagnosed through inverse participation ratios. See Fig \ref{1point05} for the slowly growing $\overline{S}$ at $\alpha = 1.05$ for different system sizes expressed as a function of $f$.

We do not give a first-principles derivation of the behavior of the EE. However, we make the observation that the $\alpha$-dependent falloff in the Hamiltonian gives rise to localization of the correlation functions. The two-point function develops a power-law decay of exponent $\beta$ with a stochastic envelope, which means for $i \neq j$:

\be
\sqrt{\overline{C^{2}_{ i j}} }  = \frac{\kappa}{d_{O}(i, j)^\beta}.   \label{emergeTI}
\ee 

In the free fermion case, the scaling of entanglement entropy is severely constrained by particle-number fluctuations \cite{Gioev_2006}, \cite{Fannes_2003}, given by $ \Delta N^2 _{A} = \Tr( C_{A} ( \mathbb I - C_{A} ) ) $ using the inequality

\be
2\Delta N^2 _{A}\leq  S(L_{A} )  \leq  \Delta N^2 _{A} O(\log(L_{A} ) ).  \label{FluctInequality}
\ee 

Using \eqref{emergeTI}, neglecting correlations and using the constraint that the ground state will generically be at half-filling, we give a non-rigorous estimate of the scaling of $\overline{S}(L_{A})$ using the behavior of $\Delta \overline{N}^2 _{A}$ in Appendix \ref{appendix:a} as a function of $\beta$, see \eqref{disorderAvgFluct}. The scaling \eqref{fractalscaling} is controlled by the exponent $\beta$. This point of view suggests that the onset of area-law behavior in disordered hopping models is linked to the the exponent $\beta$ becoming larger than  $1$. See Figure \ref{DemoBeta} for an example of $\beta$ dependence of root-mean square of the two-point functions and \ref{theorypred} for the prediction generated by  $\Delta \overline{N}^2 _{A}$.   \\

\section{Models without particle number conservation}
\label{pairing}
Here we consider translationally invariant free models of the form: 

\be
H = \sum_{\substack{i, j \\ i\neq j}} A_{ i j} c^{\dagger}_{i} c_{j} + \sum_{i, j} \frac{1}{2}( B_{i j} c^{\dagger}_{i} c^{\dagger}_{j} + h.c. ) + \mu \sum_{i} c^{\dagger}_{i} c_{i} \label{pairing_ham}
\ee
 
$B$ is antisymmetric, with entries falling off with the exponent $\alpha_p$ whereas the entries of $A$ can decay with a different exponent $\alpha_h$. Such Hamiltonians are diagonalized using a combination of Fourier and Bogoliubov transformations, or in the absence of translation invariance through the singular value decomposition of $A+B$. The ground-state reduced density matrix is characterized in terms of $\expval{c^{\dagger}_{i}c^{\dagger}_{j}}$ for  $i, j \in A$ in addition to the matrix $C_{A}$.  The formula for the EE in terms of correlation matrices is a generalization of \eqref{defnEEforC}, see \cite{Peschel_2003} for details. These models will generically contain singular excitations, though the ground state energy density is still well-defined.   
Refs    
 \cite{Vodola_2014}, \cite{Vodola_2015}, \cite{Solfanelli_2023} studied models of the form above as extensions of Kitaev's model of a superconducting wire hosting Majorana modes at the edge. For sufficiently slowly decaying couplings, these models show several exotic features such as massive edge modes, anomalous decay of correlation functions and logarithmic violations of the area law, despite a gap in the thermodynamic limit.  The latter two are related: the algebraic decay of two-point functions is a necessary but not sufficient condition for the entanglement entropy to not saturate. This point is particularly transparent for free fermion systems where Renyi entropies are directly expressed in terms of traces of powers of correlation matrices.  
 
 The exponents $\alpha_h$ and $\alpha_p$ have different roles in controlling the physics, which may be understood from considering the Hamiltonian in momentum space (after dropping an additive constant): 
 
\be
 H = \frac{1}{2} \sum^{\pi}_{k = - \pi} \begin{pmatrix} c^{\dagger}_{k} \quad c_{-k} \end{pmatrix}  \begin{pmatrix}
    a(k)+\mu   &  i b(k)  \\ 
    -i b(k) &  -a(k) - \mu 
    \end{pmatrix}  \begin{pmatrix} c_{k} \\   c^{\dagger}_{-k} \end{pmatrix}. \label{kSpacePairing_Ham}
 \ee 
 
 The functions $a(k)$ and $i b(k)$ are the symbols of the matrices $A$ and $B$ respectively, which when smooth, allow  rewriting \eqref{kSpacePairing_Ham} as an integral in the thermodynamic limit. Both $a(k)$ and $b(k)$ are $2 \pi$ periodic, but $a(k)$ is even about 0 (assuming reflection symmetry) while $b(k)$ is odd. The spectrum of single-particle excitations is given by $\lambda(k)= \sqrt{ (a(k) + \mu)^2 + b(k)^2 }$.      
 
 We examine two instances of models (\ref{pairing_ham}) in the $L \rightarrow \infty$ limit that highlight the important features in the scaling of EE:  
 \begin{equation}
  (i) \quad  A_{i j} = \frac{1}{d_{P}(i,j)^{\alpha_{h}}}  \quad \text{and} \quad B_{i j} = \frac{\text{sgn}(i-j)}{{d_{P}(i,j)^{\alpha_{p}}}} \label{LRK}
\end{equation}

In this case $a(k) = \Re \Li_{\alpha_{h}}(e^{i k} ) $ and $b(k) = \Im \Li_{\alpha_{p}}( e^{i k} ) $, where 
$\Li_{\alpha}(z)$ stands for polylogarithm of order $\alpha$. For large $\alpha_{h}$ and $ \alpha_{p}$, on tuning $\mu$ to criticality, the system falls under the $c=\frac{1}{2}$ Ising universality class. This follows from the low-energy behavior of \eqref{kSpacePairing_Ham} by taking the continuum limit, using a procedure similar to the one used in \ref{TImodel}, see \cite{Sachdev_1999} for more details. When $\alpha_{p} > 2$, for any $\alpha_{h}$, this model continues to show similar behavior to the Ising model at criticality, but shows deviations away from it. Driving $\alpha_{p} \leq 2$ alone, while keeping a local hopping term, alters the critical behavior  \cite{Vodola_2014} and results in an additive logarithmic term in the EE that persists in the presence of a gap. This additive term quickly jumps to about $\frac{1}{6} \log(L_{A})$ for $\alpha<1$. In \cite{Lepori_2016}, studying the same model, an effective field theory description of this phenomenon was developed, using modes away from the low-energy points in momentum space that remain relevant, arising from an expansion of the form: $b(k) = v(\alpha) k^{\alpha -1} + \ldots$. These terms correspond to fractional spatial derivatives and for $\alpha < 2$ may dominate low-energy behavior over the standard kinetic and mass terms. Such expansions also exist for $a(k)$, and indeed, even for dispersion of the model considered in \eqref{OBC_Ex}. However these contributions may develop divergent masses and get gapped out. We leave to the future the task of systematically studying these excitations and a field-theoretic derivation of their contribution to entanglement entropy. 

We discuss now the numerical results with $\alpha_{h} = \alpha_{p} = \alpha$, which is an instance of model studied in \cite{Vodola_2015}.           
This model can be tuned to criticality by choosing $ \mu_{c} = - a(\pi)$, in which case the spectrum of excitations becomes linear about $k = \pi$. The EE shows scaling $S(L_A) = \frac{c_{\eff}}{3} \log(L_{A})$ with $c_{\eff} = c(\alpha) + c(\mu)$. $c(\mu)$ is the critical contribution that jumps to $\frac{1}{2}$ at $\mu_{c}$ but vanishes away from it.  The $c_{\eff}$ in our notation is unrelated to the Virasoro central charge of a CFT in general. The anomalous $\alpha$-dependent contribution $c(\alpha)$ is independent of any $O(1)$ choice of $\mu$ other than $\mu_c$. It is close to $\frac{1}{2}$ as $\alpha = 0$ and approaches $0$ in a continuous manner as $\alpha$ is increased.  Notably the interacting long-range 1D antiferromagnetic Ising model displays very similar behavior of the coefficient $c_{\eff}$ as a function of $\alpha$ \cite{Vodola_2015}. 

\begin{equation}
  (i i) \quad  A_{i j} = \frac{\cos(\pi d_{P}(i,j)  )}{ d_{P}(i,j)^{\alpha}}  \quad \text{and} \quad B_{i j} = \frac{\text{sgn}(i-j)}{{d_{P}(i,j)^{\alpha}}} \label{LRKsingularityatPi}
\end{equation}

\begin{figure}[!htp]
	\centering
		\subfloat[]{\includegraphics[width=1\linewidth]{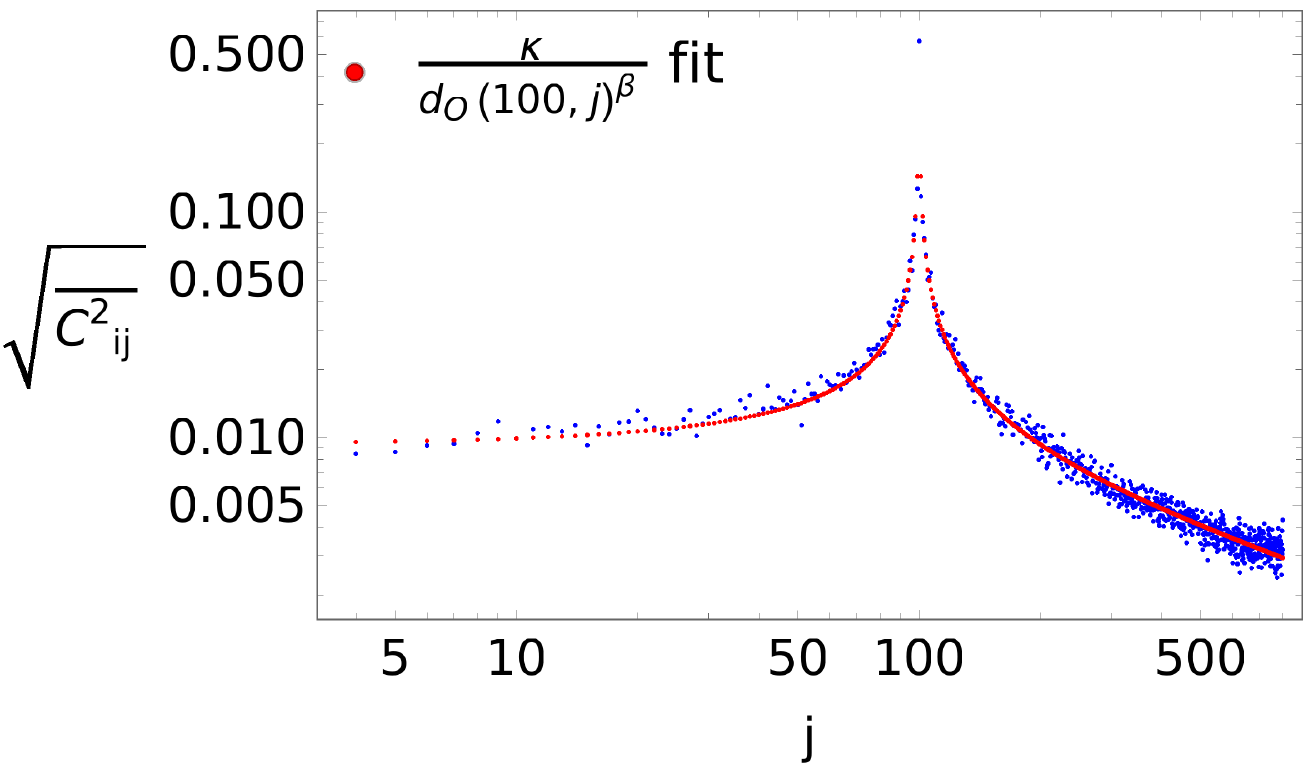}\label{DemoBeta}}
		\\
		\subfloat[]{\includegraphics[width=1\linewidth]{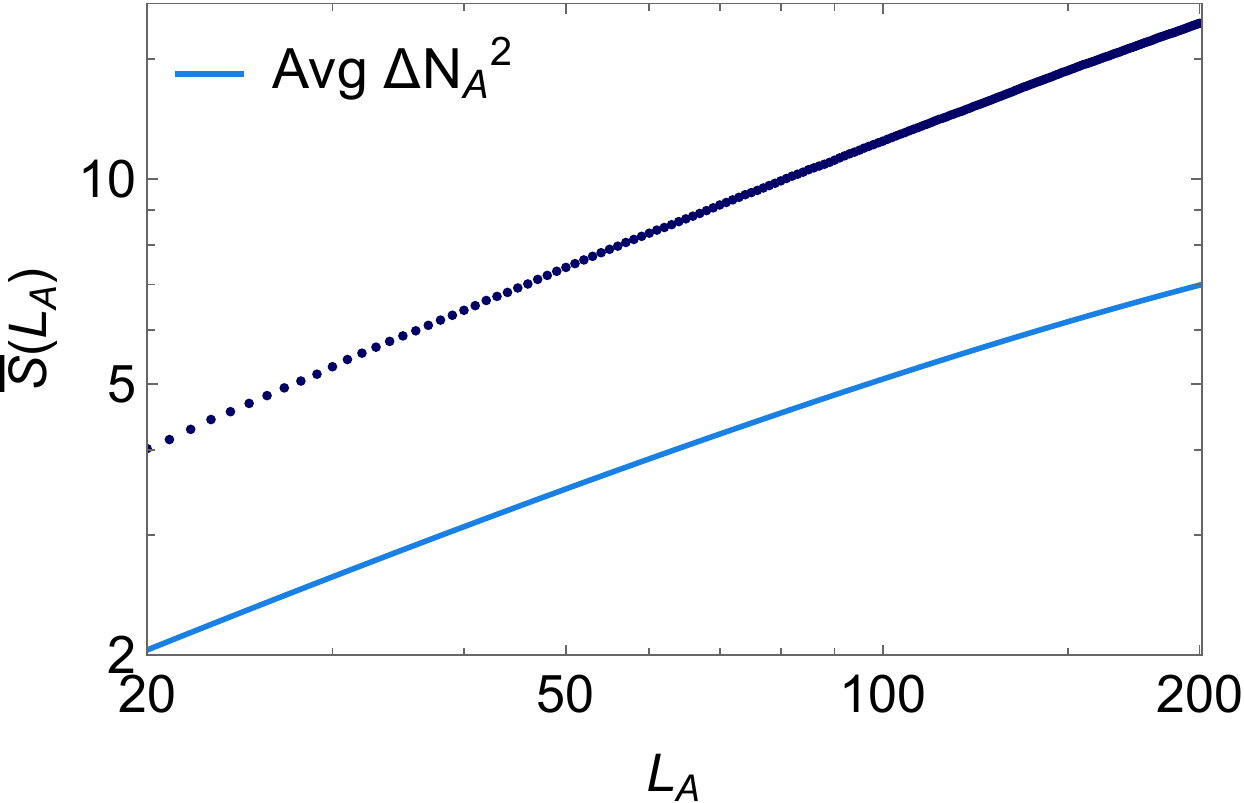}\label{theorypred}}
		\\
		\subfloat[]{\includegraphics[width=1\linewidth]{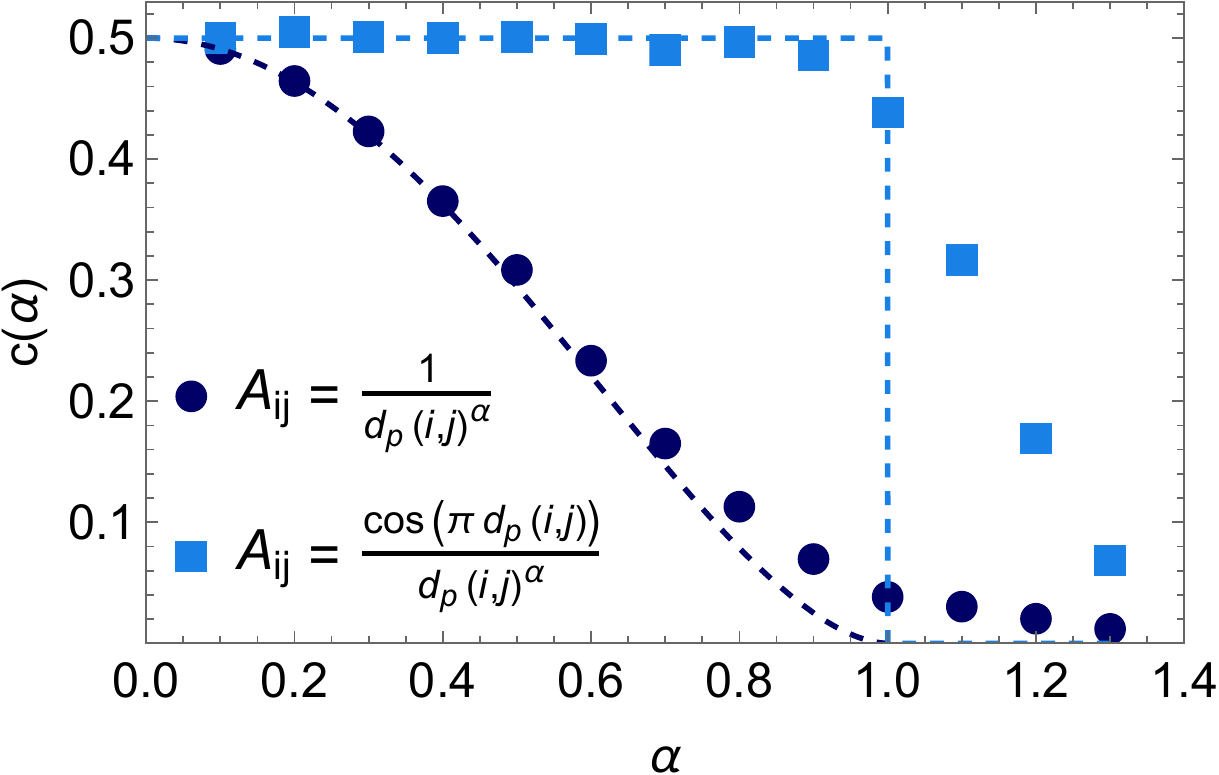}\label{ceff}}

  \caption{Fig \ref{DemoBeta} shows the root-mean square correlator $\sqrt{\overline{C^{2}_{ i j}} }$ with $i = 100$ for $L=800$ after averaging over $100$ random samples from the power-law ensemble \eqref{powerlawrm} with $\alpha=0.8$ obeying \eqref{emergeTI}. Fig \ref{theorypred} uses empirically determined parameters $m$ and $\beta$ for $L=2000$  in  \eqref{disorderAvgFluct} and makes a prediction for number fluctuations (orange line) consistent with  \eqref{FluctInequality} using \eqref{disorderAvgFluct}. $S(L_A)$ (dots) is for a system of $L=2000$ averaged over $400$ samples for the same $\alpha$. Note the agreement between the slopes for small subsystem sizes. Fig \ref{ceff} demonstrates $c_{\text{eff}}(\alpha)$ for models shown in  \eqref{LRK}, \eqref{LRKsingularityatPi}, with  $\alpha_{h} = \alpha_{p} = \alpha$ and $\mu = 0$. The superimposed dashed lines show the analytically computed $c_{\eff}$ using block Topelitz symbols.}         \label{Fig3}
  
 \end{figure}

This model of staggered hopping terms, with $\alpha = \alpha_h = \alpha_p$, has its spectrum shifted such that it cannot be tuned to a gapless point by tuning $\mu$ for $\alpha < 1$. Naturally, this model would also have a different low-energy theory. We set $\mu = 0$ and still write $ S(L_{A} ) = \frac{c_{\eff}(\alpha) }{3} \log(L_{A})$ as a convention. $c_{\eff}$ here is a function of $\alpha$ alone. For $\alpha \leq 1$, the numerically obtained $c_{\eff}$ is approximately constant and close to $ \frac{1}{2}$, and as $\alpha$ increases beyond $1$ it goes to $0$ (see Fig \ref{ceff}). 

We now interpret these numerical results in light of developments in \cite{Ares_2015}, \cite{Ares_2018} which allow the computation of the leading order terms in $S(L_A)$ in terms of the matrix symbol $G(k)$ of the block Toeplitz correlation matrix: 

\be
 C_{A}(i-j) = \begin{pmatrix} \delta_{ij} -2 \expval{ c^{\dagger}_{i} c_{j}} & 2\expval{c_{i} c_{j}}  \\ 
 2 \expval{ c^{\dagger}_{i} c^{\dagger}_{j} } & -\delta_{ij} + 2 \expval{ c^{\dagger}_{i} c_{j}}\\
 \end{pmatrix}, \label{blocktoeplitz}
\ee 

which is written in terms of the $2 \times 2$ matrix $G(k)$ as

\be  C_{A}(i-j)=\frac{1}{ 2\pi}\int_0^{2\pi} dk~G(k) e^{ik (i-j ) } \label{blocksymboldef}.\ee

Similar to the discussion below \eqref{scalarsymbol}, the asymptotic expansion of the determinant of $C_A$ was computed in \cite{Ares_2015}, \cite{Ares_2018} and it was used to determine that the leading order term in $S(L_A)$ is logarithmic with a coefficient which depends on the discontinuities of the matrix symbol $G(k)$, see Appendix \ref{appendix:c} for details. In general, the analytical value of $c_{\eff}$ has an integral representation and for \eqref{LRK} we plot it in  Fig \ref{ceff}. We find that the analytical prediction for $c_{\eff}$  vanishes at $\alpha=1$ although the numerically determined $c_{\eff}$ is nonzero and continues to decay past $\alpha=1$. For \eqref{LRKsingularityatPi} the analytical $c_{\eff}=\frac{1}{2}$ when $0<\alpha<1$ with a jump to $c_{ \eff} \approx 0.437$ at $\alpha=1$ and $c_{ \eff}=0$ for $\alpha>1$. The numerically determined $c_{ \eff}$ is close to $\frac{1}{2}$ for $\alpha < 1$ but decays gradually for $\alpha >1$. 

This apparent discrepancy between numerics and the analytical $c_{\eff}$ is resolved with the observation that the scaling of entanglement entropy is an asymptotic statement about $L_A$. In other words, in an infinite system $S(L_A)$ can continue to grow logarithmically (or otherwise) up to some point and can then transition into its true asymptotic scaling. In the long-range models we consider, even in the absence of true asymptotic logarithmic scaling $S(L_A)$ continues to grow with logarithmic behavior for finite $L_A$ until when the large $L_A$ behavior from the asymptotic expansion of Toeplitz determinants becomes valid. This is also found when one considers a theory near a critical point such that $m \ll 1$, in which one finds from field-theory calculations that $S(L_A) \sim \frac{c}{3} \log(L_A)$ for $L_A$ smaller than the scale set by $m^{-1}$ with $S(L_A)$ saturating beyond that, a scenario validated in lattice models. In case of the gapped long-range models with $\alpha>1$, the length-scale controlling the transition is not obvious. It is unclear if the logarithmic growth which saturates when $c_{\eff}$ calculated from the matrix symbol vanishes, could persist on turning on interactions. 

Our examples show that the anomalous scaling of entanglement is quite sensitive to the details of the model. The manner in which such scaling can transition to conventional scaling is non-universal. Given that the isolated divergences of $b(k)$ for $\alpha <1$ play a crucial technical role for $c_{\eff}$ to be nonzero for $\alpha>1$ as outlined in Appendix \ref{appendix:c}, it is likely that there is no violation of the area-law in the presence of a gap for $\alpha>1$ for translationally invariant models with a thermodynamic limit. That being said, there could be faster growth of entanglement for finite $L$ and for finite subsystems $L_A$ below a certain scale.

\section{Discussion}
\label{discussion}
We examined the behavior of ground-state entanglement entropy in lattice models with long-range couplings and numerically studied several examples of free fermionic models in one spatial dimension. We found a regime of intermediate fractal scaling for disordered models and for a sequence of deterministic Hamiltonians without a continuum limit, as a function of the decay exponent $\alpha$. For disordered models EE continues to be unbounded for $\alpha>1$. We provided constructions where the ground state EE approaches maximal volume-law growth consistent with the size of local Hilbert space. For systems with a continuum limit, we found that in the thermodynamic limit, the entanglement entropy is described by the predictions from the effective low-energy theory. The low-energy theory for $\alpha < 2$ may be an exotic one, featuring fractional derivative terms that give rise to algebraic decay of correlations and unbounded entanglement in the presence of a gap. In the translationally invariant models we study, the transition to conventional scaling of entanglement occurs when $\alpha>1$ and we discuss why likely $\alpha_c=1$ for this case. There is no common $\alpha_{c}$ at which conventional entanglement scaling kicks in across all free fermion systems.     

This brings us to our main result which is conceptual. Consider an infinite system. Long-range entanglement describes quantum correlations across large spatial distances and is in this sense an IR property. For translationally invariant models with a continuum limit, this implies that entanglement scaling in the IR will constrain the scaling in the UV. Here $S(L_{A})$ is precisely translationally invariant and is expected to match the predictions from UV theory at $L_{A} \ll \xi$, where $\xi$ is the length scale above which the IR scaling sets in. In $d=1$, an application of strong subadditivity  gives\footnote{Strong subadditivity applied to subsystems corresponding to intervals $l_1$, $l_2$ and the union of $l_1$, $l_2$  with separating interval of length $r$ gives $S(l_1 + r) + S(l_2 + r) \geq S(l_{1}) + S(l_{2})$ for all $l_1$, $l_2$ and $r$. We set $l_1 = l_2 = l$ and $l+r = \xi$.} $S(\xi) \geq S(l)$ for $l \ll \xi $. The inequality above constrains the range over which the UV scaling can be parametrically faster. For example, assuming $S(\xi) \sim \log(\xi )$ and $S(l) \sim l^{\gamma}$, the inequality above constrains the power-law growth to be suppressed for domains of size roughly larger than $\log(\xi)^{1/\gamma}$. Further constraints in general systems can probably be deduced on the grounds of smoothness of $S$ and dimensional analysis. There is perhaps a better formulation of these constraints by considering the putative equivalence of field-theoretic renormalization to real-space renormalization schemes like entanglement renormalization \cite{Vidal_2007}, \cite{Haegeman_2013} that proceed by removing short-distance entanglement along with rescalings.


In $d=1$, for Poincare-invariant continuum theories the constraints can be made much stronger:

\be
\frac{d}{dL_{A}} \bigg( L_{A} \frac{d S(L_{A})}{d L_{A}} \bigg) \leq 0 
\ee

implies that $S(L_A)$ cannot grow faster than $\log(L_{A})$, regardless of some nonlocal description. The concavity of $S(L_A)$ with respect to $\log(L_A)$, which follows from the Poincare invariance of $S(L_A)$, lies at the heart of the entropic $c$-theorem \cite{Casini_2007}. Conjectured generalizations of similar restrictions on scaling also exist for higher dimensions \cite{Liu_2013}.  Independent arguments based on crossovers between entanglement and thermal entropies restrict the violations of area law to be at most logarithmic \cite{Swingle_2013} assuming conventional low-temperature thermodynamics. Due to technical considerations, area laws (with potential logarithmic enhancements) are natural in continuum field theories, even nonlocal ones \cite{Solodukhin_2011}, \cite{Li_2011} although there are constructions of highly nonlocal theories with volume-law scaling \cite{Li_2011}, \cite{Shiba_2014}. The logarithmic scaling we find in the smooth long-range free fermionic systems on the lattice follows from these considerations applied to the corresponding IR continuum theories. This reasoning is consistent with results from various nonlocal lattice models \cite{Koffel_2012}, \cite{Bermudez_2017}.




Disordered systems and systems without a continuum limit are not constrained in this manner. To clarify, our statements about the continuum limit do not imply the lack of a continuum field-theoretic description of spatial or ensemble-averages for these models. Nor do our suggestions contradict the findings of entanglement scaling in local disordered models, where the logarithmic scaling obtained after a real-space RG procedure describes the entanglement up to short-distance terms \cite{Refael_2004}.   
 
It would be desirable to put our RG arguments above on a more quantitative footing. Another important task would be to carry out a systematic effective field theory calculation of entanglement entropy in the long-range lattice models discussed in our work.  As a separate matter, it would be interesting to compare the properties of nonlocal UV-complete quantum field theories with the lattice models considered here. The latter will not have Lorentz-invariant effective theories in general. Similar computations in analytically solvable $d > 1$ models may shed more light on the scaling of entanglement in reasonable nonlocal models.

\section{Acknowledgements}
We thank Alfred Shapere, Sumit Das, Ganpathy Murthy, Anatoly Dymarsky and Ahmed Khalifa for discussions and feedback. We are especially grateful to Alfred Shapere for comments on the draft. DC was partially supported by NSF  under Grant grant number PHY 2013812.   

\newpage

\appendix
\section{Fractal Scaling in Disordered Models}
\label{appendix:a} 

 The exponent $\gamma$ can be inferred from \eqref{emergeTI} as a function of $\beta$, though the derivation of $\beta$ as a function of $\alpha$ is a task to be done.  Let us assume that the density of  ground state is $\frac{1}{L} \Tr(\overline{C}) = \frac{1}{2}$. This mild assumption follows from the symmetry of averaged single-particle Hamiltonian about zero. This leads to  $\frac{1}{L} \Tr(\overline{C}^2) = \frac{1}{2}$.  Since the diagonal elements scale differently than the off-diagonal ones, let $\overline{C^2_{ii}} = m$. This fixes normalization $\kappa$ as:
 
 \be
 \kappa = \frac{ \frac{1}{2} - m }{2 (H_{2 \beta}(L-1) - \frac{H_{2 \beta-1}(L-1)}{L}  ) }
 \ee 
 
 Where $H_{j}(n)$ is the generalized Harmonic number of order $j$. Plugging in the same $\kappa$  and $m$, in expression of $\overline{N}^2 _{A}$
 \be
 \Delta \overline{N}^2 _{A} = ( \frac{1}{2} - m )( L_{A} -  \frac{  H_{2 \beta}(L_{A}-1)  L_{A} - H_{2 \beta-1}(L_{A}-1)   }{ H_{2 \beta}(L-1) - \frac{H_{2 \beta-1}(L-1)}{L}   } ) \label{disorderAvgFluct}
 \ee  

In \eqref{disorderAvgFluct} an accurate estimate is provided, consistent with the scaling of $\overline{S}(L_{A})$ on plugging in the numerically obtained $\beta$ and $m$ see Fig. \ref{theorypred}. The undetermined value of $m$ does not alter scaling of entanglement but changes the coefficient. For finite $L$ and arbitrary $\beta$ \eqref{disorderAvgFluct} does not give a simple power-law. However, for $1 < 2 \beta < 2$ and large enough  $L_{A}$ and $L$, $\overline{N}^2 _{A}$ scales as $L^{2 - 2 \beta}_{A}$. 

For the $\alpha=0$ limit, $\beta = 0$, and this computation gives the accurate estimate of $\overline{N}^2 _{A}$ as $ \frac{L_{A}}{4}  - \frac{L_{A} L }{4(L^2-L)}$.  In fact, the leading order expression for $S(L_{A}) = L_{A} \log(2)$ follows directly from analytic continuation (in $n$) of $\Tr(C_{A}^{n}) \approx L_{A}2^{-(n+1)}$, neglecting correlations among the elements of $C_{A}$  for $L_{A} \ll L$ and likewise for $( \mathbb I -C_{A} ) $. \\ 

\section{}
\label{appendix:b} 
Consider a system of even size $L$ with the interaction Hamiltonian
\be H=\sum_{i,j=1}^L V_{ij}c_i^\dagger c_j,~~~ V_{ij}=V(i-j),~~~ V(r)=\delta_{r,{L\over 2}}+\delta_{r,-{L\over 2}}.\ee
The single-particle eigenvalues are 
\be a_k=\sum_{j=1}^L V(j)e^{ij{2\pi k\over L}}=(-1)^k.\ee
The correlation matrix is given by
\be \begin{split} C(r)&={1\over L}\sum_{j=1}^L e^{2\pi i{rj\over L}}\Theta(-a_j)={1\over L}\sum_{j\text{ odd}}e^{2\pi i{rj\over L}}\\
	&={1\over 2}(\delta_{r,0}-\delta_{r,{L\over 2}}-\delta_{r,-{L\over 2}}).\end{split}\ee
For a subsystem $A$ of size $L/2$ or smaller, the correlation matrix is diagonal $C_A={1\over 2}\mathbb I$. This implies maximal growth of entanglement entropy
\be S(L_A)=L_A\log 2.\ee

\section{}
\label{appendix:c}

The correlation matrix (\ref{blocktoeplitz}) is a block Toeplitz matrix, which can be written as
\be {C_A}(i-j)={1\over 2\pi}\int_0^{2\pi} dk~G(k) e^{ik (i-j)},\ee
where the matrix symbol is
\be G(k)={1\over \lambda(k)}\begin{pmatrix}
    \alpha(k)+\mu & i b(k)\\
    -i b(k) & -\alpha(k)-\mu 
\end{pmatrix} .\ee

We apply the method of \cite{Ares_2018} to obtain $c_\text{eff}$ from this block-matrix symbol. We parametrize the latter as
\be G(k)=\cos\phi_k \sigma_z+\sin\phi_k \sigma_y .\ee

We define the coefficient of the logarithmic term to be $c_\text{eff}/3$. It is determined by the discontinuities of the symbol. Let $\{k_n\}$, be the values of $k$ at which the symbol is discontinuous. Then, $c_\text{eff}$ is given by the integral
\be {c_\text{eff}}={3\over \pi^2}\sum_n\int_{\cos\xi_n}^1 dx~\log{1-x\over 1+x}\log{\sqrt{1-x^2}\over \sqrt{x^2-\cos^2\xi_n}+\sin\xi_n},\label{intapp}\ee
where 
\be \xi_n={\delta\phi_{k_n}\over 2},\ee
and $\delta\phi_{k_n}$ is the discontinuity of $\phi_k$ at $k=k_n$.

For (\ref{LRK}), and $\alpha_h=\alpha_p=\alpha$, there is only one discontinuity at $k=0$
\be \delta\phi_0=\begin{cases}
    \pi (1-\alpha) & 0\leq \alpha<1\\
    0 & \alpha\geq 1
\end{cases}.\ee
On tuning $\mu$ to the critical value $\mu=-a(\pi)$, a second discontinuity arises at $k=\pi$, such that $\delta \phi_\pi=\pi$. In this case, the integral (\ref{intapp}) results to $c_\text{eff}={1\over 2}$.

For (\ref{LRKsingularityatPi}), and $\alpha_h=\alpha_p=\alpha$, the discontinuity is again at $k=0$
\be \delta\phi_0=\begin{cases}
    \pi & 0\leq \alpha<1\\
    2\arctan{\pi\over 2\log 2} & \alpha=1\\
    0 & \alpha> 1
\end{cases}.\ee
This leads to $c_{\text{eff}}={1\over 2}$ for $\alpha<1$ and $c_{\text{eff}}\approx 0.437$ for $\alpha=1$.

\bibliography{refs}
\end{document}